\documentstyle[aps,prl]{revtex}

\begin{document}

\draft

\widetext

\title{Time reparametrization group and the long time
 behaviour in quantum glassy systems}

\author{Malcolm P. Kennett$^{a,*}$ and Claudio Chamon$^{b,\dagger}$}

\address{ 
$^a$ Department of Physics, Princeton University, Princeton, NJ 08544
\\
$^b$ Department of Physics, Boston University, Boston, MA
02215}

\date{\today}

\twocolumn[\hsize\textwidth\columnwidth\hsize\csname@twocolumnfalse\endcsname
\maketitle

%%%%%%%%%%%%%%%%%%%%%%%%%%%%%%%%%%%%%%%%%%%%%%%%%%%%%%%%%%%%%%%%%%%%%%%%%%%%%%
\begin{abstract}
  We study the long time dynamics of a quantum version of the
  Sherrington-Kirkpatrick model. Time reparameterizations of the dynamical
  equations have a parallel with renormalization group transformations, and
  within this language the long time behaviour of this model is controlled by
  a reparameterization group (R$_p$G) fixed point of the classical dynamics.
  The irrelevance of the quantum terms in the dynamical equations in the
  aging regime explains the classical nature of the violation of the
  fluctuation-dissipation theorem.
\end{abstract}

]

%%%%%%%%%%%%%%%%%%%%%%%%%%%%%%%%%%%%%%%%%%%%%%%%%%%%%%%%%%%%%%%%%%%%%%%%%%%%%%

\narrowtext

There has been considerable interest lately in quantum systems that show
glassiness. A number of experiments \cite{Wu,Keimer,Broholm} have studied the
effect of quantum fluctuations on spin systems which are close to a quantum
phase transition between a spin glass and a quantum-disordered (paramagnetic)
phase. Theoretically, one approach to the problem has been to add quantum
fluctuations to a classical model, such as by applying a transverse field to
the Ising model \cite{Huse,DFisher}, or by giving quantum dynamics to
classical spin variables \cite{Rotor,KopUs,CugLoz}. These models are more
amenable to analytical treatment than the full $SU(2)$ quantum spin systems
\cite{BrayMoore}.

Quantum fluctuations have been shown to be responsible for redefining the
boundary between the spin glass and the quantum disordered phases in systems
such as quantum rotors or the transverse field Ising model.  In principle they
could also alter the non-equilibrium dynamics in the glassy regime at very
low temperatures. Whilst the properties in both the quantum critical and quantum
disordered regions for these systems can be understood within a static formulation \cite{Rotor},
studying the non-equilibrium effects in the glassy region requires a quantum
dynamical approach. 

In classical spin glasses the out of equilibrium nature of the system is
manifest through a persistent dependence of susceptibilities on the time
elapsed since the system entered the glassy phase. Much progress has been
achieved in recent years in understanding these aging effects in classical
spin glasses \cite{BCKM}, and only very recently have these issues begun to
be addressed in quantum systems. Cugliandolo and Lozano \cite{CugLoz} studied
a quantum version of the $p$-spin model, and in this paper we study the
long-time dynamics of the quantum version of the soft-spin
Sherrington-Kirkpatrick (SK) model \cite{SK}.

It was established in Ref. \cite{CugLoz} that quantum effects are manifest
even in the long-time ``aging'' regime of the glassy phase, contrary to the
naive expection that quantum fluctuations enter only the short-time dynamics.
However, the {\it form} of the violation of the Fluctuation Dissipation
Theorem (FDT) -- a hallmark of the aging dynamics -- coincided with that of
{\it classical} spin glass models.

One anticipates a classical limiting form for the FDT violation, because the
low frequency limit of the quantum fluctuation dissipation theorem (QFDT)
\cite{CugLoz} coincides with the $\hbar \to 0$ limit for an equilibrium
bosonic system. The aging regime is non-equilibrium; however one expects
intuitively that there is a similar situation in which quantum effects are
``irrelevant'' in modifying the form of the violation of the FDT.
 
To make these ideas more precise, we develop a general framework which allows
us to show that quantum terms in the dynamical equations are {\it irrelevant}
under a group of time reparametrization transformations that we define below.
The long-time dynamical equations, even though they contain terms purely
quantum in origin, are controlled by the classical R$_p$G {\it fixed point}
dynamical equations.  Hence, when quantum dynamics are imposed on a classical
model, the form of the violation of the FDT must coincide with that of the
classical version of the model.  Quantum effects renormalize the coefficients
in the fixed point dynamical equations, changing the details ({\it e.g.}
effective temperature) but not the form of the violation of the FDT. We make
these ideas concrete in our solution of the quantum version of the soft-spin
SK model.

Intentionally using language that parallels the Renormalization Group (RG)
terminology, let us proceed by introducing the R$_p$G transformations as a
tool to discuss the fixed point equations of the long time dynamics. 
However, R$_p$G transformations are weaker than the RG in the sense that
they act only on the correlators and dynamical equations and not on the
action.
Consider
a general transformation of the time coordinates
\begin{equation}
  \label{eq:time-trans}
  t\to \tilde t=h(t) \; ,
\end{equation}
for a differentiable function $h(t)$ with $dh/dt \ge 1$ (so as to stretch time).
Given a two-time correlation function $G(t_1,t_2)$, we define a 
transformation $G \to \tilde G$ such that
\begin{equation}
  \label{eq:G-trans}
  \tilde G(t_1,t_2) =
  \left(\frac{\partial h}{\partial t_1}\right)^{\Delta_{1}^G}
  \left(\frac{\partial h}{\partial t_2}\right)^{\Delta_{2}^G}
  G(\,h(t_1),\, h(t_2)\,) \; ,
\end{equation}
where $\Delta_{1,2}^G$ are the scaling dimensions of the correlator $G$ under
the rescaling of the time coordinates $t_{1,2}$. We term $\Delta_1^G$ and
$\Delta_2^G$ the advanced and retarded scaling dimensions respectively.

The non-equilibrium dynamics is governed by a set of integro-differential
equations for a set of correlators $\{G_\alpha\}$. We represent this set of
equations as $H[\{G_\alpha\}]=0$.

Define a fixed point of the reparametrization transformation $h(t)$ as
$H_0$ such that if the $\{G_\alpha\}$ satisfy
$H_0[\{G_\alpha\}]=0$, then $H_0[\{\tilde G_\alpha\}]=0$ is
satisfied as well.  We also write this as an equivalence $H_0 \equiv
\tilde H_0$.

Let us examine the general case when the set of equations $H$ is not a
fixed point of the reparametrization. Consider the situation when the
equations $H$ can be written as an R$_p$G invariant part $H_0$
plus a non-invariant perturbation $\delta H$:
\begin{equation}
  \label{eq:H}
  H[\{G_\alpha\}] \equiv H_0[\{G_\alpha\}]+
  \delta H[\{G_\alpha\}]=0 .
\end{equation}
The transformed equations satisfy $\tilde H \equiv H_0  
+\delta \tilde {H}$, and
$\delta H[\{\tilde G_\alpha\}]\not\equiv \delta \tilde{H} 
[\{\tilde G_\alpha\}]$, 
so $\tilde H[\{\tilde G_\alpha\}]\ne 0$.

If the transformed set of equations $\delta \tilde H [\{\tilde
G_\alpha\}] \to 0$ under repeated coordinate transformations, then the
perturbation $\delta H$ is irrelevant under time reparametrizations.
This framework allows us to determine which terms in the dynamical equations
of quantum glasses will be important in the long time limit.

We apply these ideas to a soft spin version of the SK
model \cite{inprep}. We use a closed time path (CTP) formalism
(Schwinger-Keldysh) \cite{Schwinger,Keldysh}, which obviates the need for
replicas in performing an average over quenched disorder \cite{CugLoz,CLN} for
arbitrary initial conditions. The action
$S$ has a free part $S_0$, a part due to self-interactions $S_u$, a part due
to disorder $S_{\rm dis}$, and a part due to the interaction with a thermal
bath $S_{T}$ \cite{Feynman}. We write the Lagrangians for the free part and
self interactions as

\begin{eqnarray}
{\mathcal L}_0 & = & \sum_{i,a,b} \left[ \frac{1}{2g}(\partial_t S^a_{i})
\sigma_3^{ab}
(\partial_t S^b_{i}) + \frac{1}{2}
m^2 (S^a_{i})\sigma_3^{ab}(S^b_{i}) \right]\; , \\
{\mathcal L}_u & = & \sum_{i,a,b} 
 \frac{u}{2}\left\{\left[(S^1_{i})^2\right]^2 
- \left[(S^2_{i})^2\right]^2\right\} \; ,
\end{eqnarray}
where $i,j$ are site indices and $a,b$ are Keldysh contour indices.  The
couplings are $g$, $m^2$ and $u$, and $\sigma_3$ is the Pauli matrix that acts
on the two-component fields on the time-ordered and anti-time-ordered
branches of the Keldysh contour. This quantum SK model is the $M=1$
version of the quantum rotor model considered in Ref.
\cite{Rotor}. The spin exchange interactions are infinite ranged and are
described by

\begin{equation}
{\mathcal L}_{\rm dis} = \sum_{\left<ij\right>} J_{ij}S^a_i \sigma_3^{ab}S^b_j,
\end{equation}
where the $J_{ij}$s are assumed to have a Gaussian distribution, with
$\overline{J_{ij}^2} =  J/N$, where $N$ is the number of sites and the sum
is over $i \ne j$.  We use an overbar $\overline{\cdots}$ to indicate an average
over disorder realizations and angled brackets $<\ldots>$ to indicate an
average with respect to ${\mathcal L}$.  The thermal part of the action
(obtained after integrating out heat bath variables) is

\begin{eqnarray}
S_T & = & \int_0^\infty dt_1 \int_0^\infty dt_2 \sum_{i,a,b}
 S^a_i(t_1)\left\{i\nu(t_1 - t_2)[\delta^{ab} 
- \sigma_1^{ab}] \right. \nonumber \\
& & \left.- \eta(t_1 - t_2)[\sigma_1^{ab} + 
i\sigma_2^{ab}]\right\}S^b_i(t_2) ,
\end{eqnarray}
where $\eta$ and $\nu$ are the same heat bath kernels for noise and
dissipation that are described in Ref. \cite{CugLoz}.  We write the
generating functional for the action in the CTP formalism and 
average over disorder then decouple the four spin term with a
Hubbard-Stratonovich transformation.

We define the correlation $C(t_1,t_2)$ and response $R(t_1,t_2)$ for times $t_1$
and $t_2$ after the initial quench as

\begin{equation}
C_{ij}(t_1,t_2) = \frac{1}{2}\overline{\left<S^1_i(t_1)S^1_j(t_2) +
    S^2_i(t_1)S^2_j(t_2) \right>},  
\end{equation}
and $R_{ij}(t_1,t_2) = \delta/\delta h_j(t_2)
\overline{\left<S_i^1(t_1)\right>}$, which in linear response theory may be
written as 
\begin{equation}
R_{ij}(t_1,t_2) =
\frac{i}{\hbar}\overline{\left<S^1_i(t_1)[S^1_j(t_2)-S^2_j(t_2)]\right>} \; .
\end{equation}
We use a saddle point evaluation ($N \to \infty$) and expand to $O(u^2)$ to
obtain mode coupling equations for $C$ and $R$.  We solve these equations in
the stationary regime for the paramagnetic phase and the spin glass phase.

In the paramagnetic phase, the correlation functions are time translation 
invariant (TTI) once initial transients have decayed and we can
Fourier transform them.  The heat bath kernels $\eta$ and $\nu$ satisfy the
QFDT \cite{CugLoz}
if there is Ohmic dissipation, and this implies that the same is true
for the correlation and response in the paramagnetic phase
\begin{equation}
C(\omega) =
\frac{\hbar}{2}\coth\left(\frac{\beta\hbar\omega}{2}\right)\left[R(\omega) -
 A(\omega)\right],
\end{equation}
where $A(\omega)$ is the Fourier transform of the advanced response ($R$ is
the retarded response) and $\beta$ is the inverse temperature.  The
correlators may be written as the sum of a stationary (TTI)
piece and an aging piece
\begin{eqnarray}
R(t_1,t_2) & = & R_{ST}(t_1 - t_2) + R_{AG}(t_1,t_2) \; , \\
C(t_1,t_2) & = & C_{ST}(t_1 - t_2) + C_{AG}(t_1,t_2) \; .
\end{eqnarray}
The stationary pieces decay on a much shorter timescale than the aging
pieces, so only the aging pieces influence the long time dynamics.  In the
spin glass phase, we need to take into account the aging that has occurred up
to the waiting time, and we find the expected zero frequency piece in the
response, so that

\begin{eqnarray}
R_{ST}(\omega) - A_{ST}(\omega)  & =&\frac{2}{\hbar} 
\tanh \left( \frac{\beta\hbar\omega}{2} \right) C_{ST}(\omega) \nonumber \\
& & + \frac{4\pi\beta J^2}{(M\gamma_0)} q_{EA} 
\left(\chi_\infty + (RC)_\infty \right) \delta(\omega) \; , 
\end{eqnarray}
where $q_{EA}$ is the Edwards-Anderson order parameter, $M\gamma_0$ is the
friction coefficient for the bath, $\chi_\infty = \lim_{t_1 \to \infty}
\int_0^{t_1} dt \, R_{ST}(t_1 - t)$, and $(RC)_\infty$ is (noting that $t_1
\sim t_2 \gg |t_1 - t_2|$ in the evaluation of the integral)

\begin{equation}
(RC)_\infty = \lim_{t_1 \to \infty} \int_0^{t_1} dt \, R_{AG}(t_1,t)\left(C_{AG}(t,t_1)
 + C_{AG}(t_1,t)\right) \; .
\end{equation}
We implement weak ergodicity breaking (WEB) and weak long term memory (WLTM)
\cite{Bouchaud,CugKurA} in the integrals of correlation functions as
described in Ref. \cite{CugLoz} to obtain the dynamical equations in the
aging regime for which $t_1 > t_2$ and both $t_1$ and $t_2$ become very large

\begin{eqnarray}
  \label{eq:mode-coupling_one}
  H^R[C,R] &=& H_0^R[C,R] + \delta H^R[C,R] =0 \; , \\
  \label{eq:mode-coupling_two}
  H^C[C,R] &=& H_0^C[C,R] + \delta H^C[C,R] =0 \; ,
\end{eqnarray}
where

\begin{eqnarray}
\label{eq:H_0^R}
H^R_0[C,R] & = & \lambda_1 R(t_1,t_2) + \lambda_2 C(t_1,t_2)^2 R(t_1,t_2)
\nonumber \\ & &
+ 2J^2\int_0^\infty dt\,  R(t_1,t)R(t,t_2) \nonumber \\
& & + \lambda_3 \int_0^\infty dt\,
C(t_1,t)^2 R(t_1,t)R(t,t_2) \; , \\
\label{eq:deltaHR}
\delta H^R[C,R] & = & -\frac{1}{2g} \frac{\partial^2}{\partial t_1^2} 
R(t_1,t_2 )- \frac{1}{3}\lambda_2 \hbar^2 R(t_1,t_2)^3 
\nonumber \\ 
& & - \frac{1}{3}\lambda_3 \hbar^2 \int_0^\infty dt\, R(t_1,t)^3 R(t,t_2) \; ,
\end{eqnarray}
and
\begin{eqnarray}
\label{eq:H_0^C}
H^C_0[C,R] & = & \lambda_1 C(t_1,t_2)   
+ \frac{1}{3}\lambda_2 C(t_1,t_2)^3   \nonumber \\
& & + 2J^2\int_0^\infty dt\left[C(t_1,t)R(t_2,t) + R(t_1,t)C(t,t_2)\right] \nonumber \\
& & + \frac{1}{3}\lambda_3\int_0^\infty dt\, C(t_1,t)^3 R(t_2,t) \nonumber \\
& & + 
\lambda_3 \int_0^\infty dt\, C(t_1,t)^2 R(t_1,t) C(t,t_2) \; , \\
\label{eq:deltaHC}
\delta H^C[C,R] & = &  -\frac{1}{2g} \frac{\partial^2}{\partial t_1^2} 
C(t_1,t_2) + \lambda_4 R(t_1,t_2) 
\nonumber \\ 
& & + \lambda_5 C(t_1,t_2)^2 R(t_1,t_2)
-\frac{1}{3}\lambda_5 \hbar^2  R(t_1,t_2)^3 \nonumber \\
& &
-\frac{1}{9}\lambda_2 \hbar^2 C(t_1,t_2)R(t_1,t_2)^2  \nonumber \\
& & -\frac{1}{3}\lambda_3 \hbar^2 \int_0^\infty dt 
\left[R(t_1,t)^3 C(t,t_2) \right. 
 \\ 
& & \left. \hspace*{1cm} + 3 R(t_1,t)^2 R(t_2,t)C(t_1,t)\right] \; . \nonumber 
\end{eqnarray}
The coefficients $\lambda_1,\ldots,\lambda_5$ depend on $\hbar$, $u$, $J$, as
well as temperature and couplings to the bath.

Let us now consider the time coordinate transformation defined in Eq.
(\ref{eq:time-trans}). Repeated application of R$_p$G transformations leads
to the fixed point equations $H_0^R[C,R] =0$ and $H_0^C[C,R] =0$, if the
$C,R$ correlators are chosen to transform under Eq.  (\ref{eq:G-trans}) with
dimensions $\Delta^C_{1,2}=0$, $\Delta^R_1=0$, and $\Delta^R_2=1$ (suggested by 
the form of the FDT):
\begin{eqnarray}
  \label{eq:CR-trans}
  \tilde C(t_1,t_2) &=&
  C(\,h(t_1),\, h(t_2)\,) \; ,\\
  \tilde R(t_1,t_2) &=&
  \left(\frac{\partial h}{\partial t_2}\right)
  R(\,h(t_1),\, h(t_2)\,) \; .
\end{eqnarray}
The fixed point equations contain terms of the same form as considered in
Ref. \cite{CugKurB} in the dynamical solution of the {\it classical} SK
model.  These equations have a solution via the ansatz
\begin{equation}
\label{eq:modified_FDT} 
R(t_1,t_2) = \beta X[C(t_1,t_2)]\frac{\partial}{\partial t_2}C(t_1,t_2)
\theta(t_1 - t_2),
\end{equation}
which is invariant under time reparametrization.  Equation 
(\ref{eq:modified_FDT}) is the modified version of the FDT (in the FDT, 
$X = 1$).  The term $X[C]$ is the FDT violation factor, which we calculate 
here for the model in question.

The quantum terms are
irrelevant under the R$_p$G transformations, as are some of the classical
terms.  Hence, classical dynamics is recovered in the long time limit even
when $\hbar \ne 0$.  In Ref. \cite{CugKurB} the asymptotic equations were
obtained by ignoring terms such as time derivatives of the correlation and
response, whereas here these terms are seen to be irrelevant under repeated
application of a R$_p$G transformation.

A simple example which illustrates the idea of irrelevant terms, is 
to consider the reparametrization $h(t) = bt$, with $b>1$.  If
we apply this transformation to the $R^3$ term in $\delta H^C$ it is rescaled
by $b^{-3}$, whereas $H_0^C \rightarrow H_0^C$.  This term will clearly
become negligible under repeated transformations.

We find the solution to equations (\ref{eq:mode-coupling_one}) and
(\ref{eq:mode-coupling_two}) with the ansatz (\ref{eq:modified_FDT}) to be
similar to that found by \cite{CugKurB}
\begin{equation}
X[C] = \frac{\lambda_2 C}{\beta} 
\frac{(2J^2 + \lambda_3 q_{EA}^2)^{\frac{1}{2}}}{(2J^2 +
 \lambda_3 C^2)^{\frac{3}{2}}},
\end{equation}
where for small $q_{EA}$, {\it i.e.} $\lambda_3 C^2 < \lambda_3 q_{EA}^2 \ll
2J^2$, we recover $X[C]= \lambda_2 C/(2J^2\beta)$. 

In the solution above, $\lambda_3 = \frac{3}{2}u^2$, and quantum effects can
only enter through the coupling $\lambda_2=\frac{3}{2}u^2 \chi_\infty .$
Hence, in the fixed point dynamical equations (\ref{eq:H_0^R}) and
(\ref{eq:H_0^C}), quantum effects enter only as renormalizations of the 
coefficients.

The picture we have of the dynamics of a quantum spin glass is that at short
times the QFDT holds with a zero frequency correction, then at intermediate
times, quantum fluctuations can still affect the dynamics in a significant
way \cite{CugLoz}, but at longer times they become increasingly irrelevant.
Finally, at very long times, $R$ and $C$ are related via the classical R$_p$G
fixed point equation, although the matching conditions between this long time
behaviour and the short time behaviour are also affected by quantum effects.
It is in this sense that we call the long time dynamical equations
with renormalized coefficients {\it renormalized classical}.

In the $\hbar \to 0$ limit, the CTP formalism
becomes identical to the Martin-Siggia-Rose (MSR) \cite{MSR} functional
integral formalism \cite{CugLoz}.  However when $\hbar \ne 0$ the dynamics in each approach
enter in very different ways -- in the MSR approach, the dynamics are imposed
on the system through a Langevin equation describing the interaction with the
bath, but in the CTP formalism, the dynamics are intrinsic to the quantum
spin variables.  The reason that both approaches give the same long time
dynamics is that the time derivatives and bath interactions are irrelevant
under R$_p$G transformations in either case.

The time reparametrization invariance of the long time dynamics
of the classical SK model was noticed by several authors
\cite{CugKurB,Everybody}. The quantum dynamical contribution, as well as
other classical and derivative terms that we discuss in this paper break that
invariance. However, in the language of the R$_p$G, these terms are
irrelevant. This suggests that R$_p$G invariance can be a guiding principle
in understanding the long time behaviour of glassy systems. In particular,
R$_p$G invariance could be the symmetry principle related to the modified
FDT; the retarded and advanced scaling dimensions of $C,R$ are tied by the
form of the modified FDT, and {\it vice versa}, the form of the modified FDT
is constrained by the scaling dimensions of the correlators $C,R$. Along
these lines, it would be very interesting to look at different models that
could lead to mode-coupling equations with alternative R$_p$G fixed points.
These could possibly have more than two correlation functions, and could
contain correlators with different scaling dimensions than in the usual
classical models (and their quantum descendents). Most likely candidates are
models with non-trivial spin algebras, such as the SU(2) spins for which a
dynamical approach has very recently been developed \cite{Kiselev}, and the
$SU(N)$ models \cite{SUN}.

%%%%%%%%%%%%%%%%%%%%%%%%%%%%%%%%%%%%%%%%%%%%%%%%%%%%%%%%%%%%%%%%%%%%%%%%%%%%%%

We would like to thank Jinwu Ye for useful discussions and Leticia Cugliandolo 
and David Huse for clarifying discussions and careful reading of the manuscript.
We also acknowledge the hospitality of the Institute for Advanced Study at
Princeton (C.~C), where this work was started, and of Boston University
(M.~P.~K), where this work was completed. Support was provided by the NSF
Grant DMR-98-76208, and the Alfred P. Sloan Foundation (C.~C).

%%%%%%%%%%%%%%%%%%%%%%%%%%%%%%%%%%%%%%%%%%%%%%%%%%%%%%%%%%%%%%%%%%%%%%%%%%%%%%
\noindent * Electronic address: mkennett@princeton.edu \\
$\dagger$ Electronic address chamon@bu.edu

\end{document}